\documentclass[fleqn,usenatbib]{mnras}

\usepackage{newtxtext,newtxmath}

\usepackage[T1]{fontenc}

\DeclareRobustCommand{\VAN}[3]{#2}
\let\VANthebibliography\thebibliography
\def\thebibliography{\DeclareRobustCommand{\VAN}[3]{##3}\VANthebibliography}

\usepackage{graphicx}	
\usepackage{amsmath}	

\newcommand\AJ{AJ}

\newcommand\psj{Planet. Sci. J.}

\title[SOAR 3I/ATLAS Observations]{Near-Discovery SOAR Photometry of the Third Interstellar Object: 3I/ATLAS}

\author[Frincke et al.]{
Tessa T. Frincke,$^{1}$\thanks{E-mail: frincket@msu.edu}
Atsuhiro Yaginuma,$^{1}$
John W.\ Noonan,$^{2}$
Henry H.\ Hsieh,$^{3}$
Darryl Z. Seligman, $^{1,4}$
\newauthor
Carrie E.\ Holt,$^{5}$
Jay Strader, $^{1}$
Thomas Do,$^{1}$
Peter Craig,$^{1}$
Isabella Molina$^{1}$
\\
$^{1}$Department of Physics and Astronomy, Michigan State University, East Lansing, MI 48824, USA\\
$^{2}$Department of Physics, Auburn University, Edmund C.\ Leach Science Center, Auburn, 36849, AL, USA\\
$^{3}$Planetary Science Institute, 1700 East Fort Lowell Rd., Suite 106, Tucson, AZ 85719, USA\\
$^{4}$NSF Astronomy and Astrophysics Postdoctoral Fellow\\
$^{5}$Las Cumbres Observatory, 6740 Cortona Drive, Suite 102, Goleta, CA 93117, USA
}

\date{Accepted XXX. Received YYY; in original form ZZZ}

\pubyear{2025}

\begin{document}
\label{firstpage}
\pagerange{\pageref{firstpage}--\pageref{lastpage}}
\maketitle

\begin{abstract}
3I/ATLAS was discovered on UT 2025 July 1 and joins a limited but growing population of detected $\sim10^2-10^3$ m scale interstellar objects. In this paper we report photometric observations of 3I/ATLAS from the nights of UT 2025 July 3, UT 2025 July 9, and UT 2025 July 10 obtained with the Southern Astrophysical Research Telescope (SOAR). The photometric observations are taken with the Goodman High Throughput Spectrograph (HTS) in the $r'$-band. These data provide 28 photometric data points to the rapidly growing composite light curve of 3I/ATLAS. They reveal that the object did not exhibit obvious long-term variability in its brightness when these observations were taken. These observations appear to have captured two moderate and independent brightening events on UT 2025 July 9, and UT 2025 July 10. However, we perform a series of stellar contamination, stacking, and aperture experiments that demonstrate that the increases in brightness by $\sim0.8$ magnitudes appear to be a result of poor seeing and stellar contamination by close-proximity field stars. We report the mean brightnesses of 3I/ATLAS on each night of magnitude 18.14, 17.55, and 17.54 for UT 2025 July 3, 9, and 10, respectively. Moreover, the presence of cometary activity in extant images obtained contemporaneously with these data precludes them from revealing insights into the rotation of the nucleus. We conclude that the activity of 3I/ATLAS on UT 2025 July 9 and UT July 10 was consistent with the near-discovery activity levels, with no obvious outburst activity.
\end{abstract}


\section{Introduction} \label{sec:intro}

The Asteroid Terrestrial-impact Last Alert System (ATLAS) survey \citep{Tonry2018a} discovered a third macroscopic interstellar object in the Solar System on UT 2025 July 1 \citep{Denneau2025}. This object was originally named A11pl3Z, and was later renamed 3I/ATLAS (C/2025 N1) when its interstellar nature was confirmed. The object was discovered when it was between the orbits of Mars and Jupiter in the outer asteroid belt at a heliocentric distance of $\sim4.2$ au. 

The first two interstellar objects 1I/`Oumuamua and 2I/Borisov were discovered in 2017 \citep{Williams17} and 2019 \citep{borisov_2I_cbet}, respectively. These objects displayed divergent physical properties, yet all around 2I/Borisov was more similar to 3I/ATLAS. 1I/`Oumuamua exhibited brightness variations by a factor of $\sim 12$ corresponding to an oblate spheroid best fit shape with dimensions of $6:6:1$ \citep{Mashchenko2019,Taylor2023} based on the full composite light curve presented by \citet{Belton2018} with observations reported by \citet{Meech2017,Jewitt2017,Ye2017,Fraser2017,Knight2017}. 1I/`Oumuamua also exhibited a strong and radial nongravitational acceleration on its trajectory as reported by \citet{Micheli2018} but no dust coma. Limits on the dust production of 1I/`Oumuamua were placed by \citet{Meech2017,Jewitt2017,Ye2017}. Moreover,  limits on the gas production of 1I/`Oumuamua were set for variety of species, including carbon based species such as CO and CO$_2$ from the \textit{Spitzer Space Telescope} \citep{Trilling2018}. Moreover, its reflectance spectra was measured as moderately red \citep{Meech2017,Fitzsimmons2018,Ye2017,Masiero2017} 2I/Borisov displayed typical signs of cometary activity \citep{Jewitt2019b,Fitzsimmons:2019,McKay2020,Guzik:2020,Hui2020,Kim2020,Cremonese2020,yang2021,Opitom:2019-borisov, Kareta:2019, lin2020,Bannister2020,Xing2020,Bagnulo2021,Aravind2021,Deam2025} and was enriched in CO compared to H$_2$O \citep{Bodewits2020, Cordiner2020}. For reviews that are up to date to the point before 3I/ATLAS was discovered see \citet{Jewitt2023ARAA,MoroMartin2022,Fitzsimmons2024,Seligman2023}.

Post-discovery observations of 3I/ATLAS revealed that it was active, reddened, and had a mostly flat light curve. \citet{Jewitt2025} reported observations with the Nordic Optical Telescope on UT 2025 July 2 that showed that 3I/ATLAS was active. \citet{Alarcon2025} also reported observations on the night of UT 2025 July 2 from the Two-meter Twin Telescope (TTT3) at Teide Observatory that confirmed the presence of cometary activity. As additional evidence of sublimative driven mass-loss, Figure 3 in \citet{Seligman2025} showed cometary activity in  deep stacks of images from UT 2025 July 2 and 4 from the ESO's Very Large Telescope (VLT) and the Canada France Hawaii Telescope (CFHT).

Precovery observations of 3I/ATLAS have revealed that the activity identified close to the discovery date was in fact sustained for significant amount of time prior to the discovery when the object was at 4.2 au. For example, \citet{Chandler2025} reported precovery images dating back to UT 2025 June 21 with the Scientific Validation (SV) images of the Rubin Observatory. These revealed that cometary activity was present as early as June 2025 when the object was at 4.85 au. \citet{Feinstein2025} reported precovery observations with the Transiting Exoplanet Science Satelite \textit{TESS}, demonstrated that the brightness of 3I/ATLAS was constant from UT 2025 May 2 when 3I/ATLAS was at 6.2 au suggesting hypervolatile activity. A contemporaneous and independent analysis of the \textit{TESS} images agree with these results \citep{Martinez-Palomera2025}.

The current orbital solution for 3I/ATLAS shows noticeable deviation from the orbital properties of the first two interstellar objects. 3I/ATLAS had a relatively fast excess velocity at infinity of $v_\infty \sim58$ km s$^{-1}$, significantly faster than 1I/`Oumuamua and 2I/Borisov. This indicates that the object was relatively old compared to dynamical ages inferred of the first two interstellar objects \citep{Mamajek2017,Gaidos2017a, Feng2018,Fernandes2018,Hallatt2020,Hsieh2021}. Moreover, \citet{Taylor2025} and \citet{Hopkins2025b} argued that 3I/ATLAS likely formed around a low metallicity star (${\rm [Fe/H]} \le -0.04$), and that its existence along with that for 1I/`Oumuamua and 2I/Borisov - implies that interstellar object formation was efficient at early galactic ages. 

In addition to imaging, spectral data has provided evidence to support claims of cometary activity. \citet{Yang2025} reported optical and near-infrared spectroscopic observations of 3I/ATLAS obtained with the Gemini South Telescope with the Gemini Multi-Object Spectrograph (GMOS) on UT 2025 July 4 and 5. They also reported observations obtained with the SpeX spectrograph on the NASA Infrared Telescope Facility (IRTF) and UT 2025 July 14. SOAR spectroscopic data were also obtained UT 2025 July 4 covering 370--700 nm which showed no detectable gas emission with a notable coma and suggested an alternative mechanism for coma formation \citep{Puzia2025}. Ultraviolet images obtained with the \textit{Neil Gehrels-Swift} Observatory from UT 2025 July 31 to August 1 offer the first indication of water activity via the detection of an OH emission feature \citep{Xing2025}. Observations from the \textit{SphereX} spacecraft \citep{Lisse2025} and \textit{JWST} \citep{Cordiner2025} revealed a high CO$_2$ production, and high CO$_2$ to H$_2$O ratio. \citet{Rahatgaonkar2025} measured Ni vapor and CN production from VLT observations. 
 
Numerous spectroscopic observations of 3I/ATLAS have been reported which confirm early claims of a red spectral slope. \citet{Opitom2025} reported spectral observations from UT 2025 July 3 with the MUSE instrument on VLT that confirmed the reddened reflectance spectra and confirmed the cometary nature. \citet{Belyakov2025} reported spectral measurements from 420--1000 nm taken with the Palomar 200-inch telescope and Apache Point Observatory. They reported non-detections of C$_2$ emissions along with a consistently reddened reflectance spectrum. Near-discovery observations of 3I/ATLAS with the NASA IRTF showed a similarly reddened color and spectral slope \citep{Kareta2025}. 

\begin{table*}

    \centering
    \caption{Goodman HTS  Red Camera Observation Log. Phase angle, heliocentric, and geocentric distances all queried directly from JPL Horizons. We report the average seeing that was measured for a given set of exposures. The asterisk indicates that the seeing was not measured for all or a portion of the exposures because the Ring-Image Next Generation Scintillation Sensor (RINGSS) was down \citep{Tokovinin2021}.}
    \begin{tabular}{cccccccc}
     Date&Observation&  Filter&  Exposure& Average Seeing&Heliocentric &Geocentric &Phase \\
      (UT)& & & (s)&(arcsec) & Distance (au)& Distance (au)&Angle (degree)\\
      \hline
    2025-07-03&Imaging &  r-SDSS&  8 $\times$ 90.0& *& 4.44 & 3.44 & 2.637\\
    2025-07-09&Imaging& i-SDSS& 15 $\times$ 30.0 &0.575$^{*}$& 4.24 & 3.27 & 4.680\\
    . . .&. . .& r-SDSS& 50 $\times$ 30.0 &0.620$^{*}$& . . .& . . .&. . .\\
    2025-07-10& Imaging& i-SDSS&5 $\times$ 30.0 & 0.770& 4.20 & 3.24 & 5.049\\
     . . .& . . .& g-SDSS&7 $\times$ 30.0 &0.796$^{*}$& . . .& . . .&. . .\\
     . . .& . . .& r-SDSS&105 $\times$ 30.0 & *& . . .& . . .&. . .\\ 
    \end{tabular}

    \label{tab:obs_log}

\end{table*}

Recent photometric observations indicate that 3I/ATLAS has minimal brightness variation and color consistent with the first two interstellar objects. \citet{Marcos2025} reported the potential identification of a $P_{\rm rot}=16.79$ h rotational period from time-series photometry obtained with the OSIRIS camera spectrograph at the 10.4-meter Gran Telescopio Canarias and the Two-meter Twin Telescope. \citet{Santana-Ros2025} similarly reported a nucleus spin rate with rotational period of $P_{\rm rot}=16.16\pm0.01$ h with a light curve amplitude of $\Delta m=0.3$ mag based on spectral and photometric observations with the SALT telescope and Nordic Optical Telescope. \citet{Seligman2025} reported additional photometric observations with ATLAS, the Faulkes Telescope North and South, LCO, Telescope Joan Oro, and TRAPPIST North and South of 3I/ATLAS. They found a light curve that was mostly flat (within the uncertainties stemming from combining data from various telescopes and reduction techniques) from UT 2025 July 2--5. They reported colors that were red similar to D-type asteroids, 1I/`Oumuamua, and 2I/Borisov. Photometric data obtained by \citet{Beniyama2025} on UT 2025 July 15 also indicate a red surface color and no significant light curve variability. Using photometric data from the Hubble Space Telescope (HST), \citet{Jewitt2025_HST} confirmed cometary activity and provided an upper limit to the nuclear radius of $r_n<2.4$ km based on the surface brightness profile. There is an absence of observational data reported in literature thus far in the window of UT 2025 July 9 and 10. This work fills this gap and adds to an expanding composite light curve with photometric measurements of 3I/ATLAS. Additional observations of 3I/ATLAS have been reported contemporaneously with those reported here \citep{SalazarManzano2025, Gray2025, Tonry2025, Ye2025, Hutsemekers2025, Coulson2025, Hoogendam2025, Jewitt2025_NOT}, including theoretical estimates \citep{Guo2025, Perez-Couto2025, Grant2025, Keto2025}.

In this paper, we present data obtained on UT 2025 July 3, 9 and 10 with the Southern Astrophysical Research Telescope (SOAR) of 3I/ATLAS. We review the data obtained, the data reduction, and photometric analysis methodology in Section \ref{sec:obs}. We present the composite light curve results  in Section \ref{sec:comp_lc} and conclude in Section \ref{sec: discussion}.

\begin{figure}
    \centering
    \includegraphics[width=1.\linewidth]{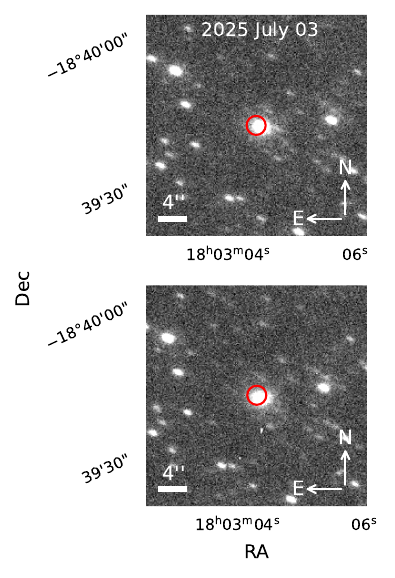}
    \caption{Reduced and rotated images from observations on the SOAR telescope with the Goodman HTS instrument of 3I/ATLAS on UT 2025 July 3. Each panel is a single 90 s exposure taken in r'-band and 3I/ATLAS is identified within the red circle. Arrows indicate directions of North (N) and East (E). Associated animation is available online. }
    \label{fig:0702_obs}
\end{figure}

\section{SOAR Observations of 3I/ATLAS} \label{sec:obs}

\subsection{Goodman HTS Imaging}\label{subsec:goodman}

\begin{figure}
    \centering
    \includegraphics[width=1.\linewidth]{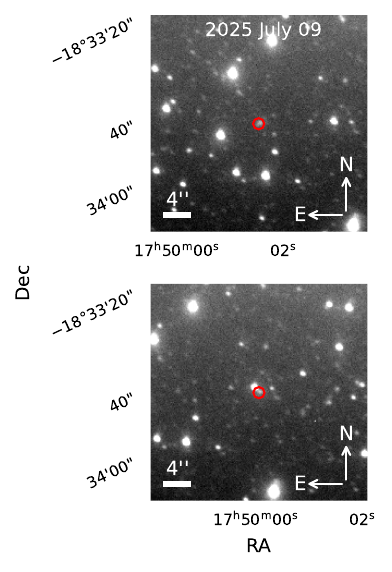}
    \caption{Reduced and rotated images from observations of 3I/ATLAS obtained with  the SOAR/Goodman HTS on UT 2025 July 9. Each panel is a single 30 s exposure taken in $r'$-band and 3I/ATLAS is identified within the red circle. Arrows indicate directions of North (N) and East (E).  Associated animation is available online.}
    \label{fig:0709_obs}
\end{figure}

\begin{figure}
    \centering
    \includegraphics[width=1.\linewidth]{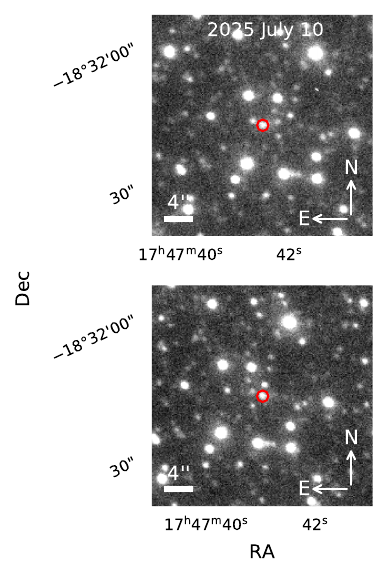}
    \caption{Reduced and rotated images from observations on the SOAR/Goodman HTS of 3I/ATLAS on UT 2025 July 10. Each panel is a single 30 s exposure taken in $r'$-band and 3I/ATLAS is identified within the red circle. Arrows indicate directions of North (N) and East (E).  Associated animation is available online.} 
    \label{fig:0710_obs}
\end{figure}

Observations of 3I/ATLAS were conducted with the 4.1-meter Southern Astrophysical Research (SOAR) telescope on Cerro Pach\'on in Chile. The observations of the third interstellar object presented in this paper were acquired on UT 2025 July 3, 9, and 10. These data were acquired with the imaging mode of the Goodman High Throughput Spectrograph \citep{Clemens2004}, with the Goodman Red Camera. The images were binned 2 $\times$ 2, giving a pixel scale of $0\farcs3$~pixel$^{-1}$ over a circular $7\farcm5$ field of view. 

\begin{figure}
    \centering
    \includegraphics[width=1.\linewidth]{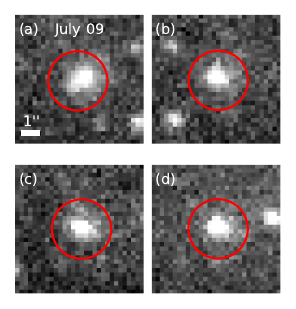}
    \caption{Series of zoomed in images from observations of 3I/ATLAS on UT 2025 July 9. Each panel is a single 30.0 s exposure taken in $r'$-band and 3I/ATLAS is identified within the red circle.}
    \label{fig:0709_grid}
\end{figure}

On UT 2025 July 3, images in the $r'$-band were acquired using a 90 s exposures. These observations are shown in Figure \ref{fig:0702_obs} and exhibit some trailing due to the longer 90 s exposure time, which causes the red circle highlighting the position of 3I/ATLAS to appear off center. On UT 2025 July 9, observations were focused on imaging of 3I/ATLAS in the Sloan $r'$-band in 30 s exposures spanning over $\sim 25$ minutes with some $i'$- band taken in 30 s exposures. Observations conducted on UT 2025 July 10 included images in the $g'$-, $r'$-, and $i'$- bands with a continued emphasis on imaging in the $r'$-band in 30 s exposures over $\sim 50$ minutes. Reduced $r'$-band frames from July 9 and 10 are shown in Figures \ref{fig:0709_obs} and \ref{fig:0710_obs}, respectively. A detailed summary of all observations is provided in Table \ref{tab:obs_log}. In Figure \ref{fig:0709_grid} we show a series of zoomed in images of 3I/ATLAS from the night of UT 2025 July 9,  which had the best overall seeing. The target was observable for the entirety of each observing window on UT 2025 July 3, 9, and 10 with airmass less than 2 and relatively clear weather conditions. On UT 2025 July 9 and 10, there was significant moonlight contribution to the overall sky brightness as a result of the full moon. The average moon illumination percentage for each observation as reported by JPL Horizons is 53.47\%, 97.00\%, and 99.42\%, respectively.

\subsection{Imaging Data Reduction} \label{subsec:data_reduction}

Raw images from the Goodman HTS were processed 
with a procedure adapted from the Astropy CCD Data Reduction Guide \footnote{\url{http://www.astropy.org/ccd-reduction-and-photometry-guide/}}. In addition to the raw science images, calibration images including bias frames and flatfield images from observations of the illuminated interior of the closed dome as well as of the twilight sky were captured on UT 2025 July 3 and 9 to compensate for variations in voltage, sensitivity, illumination, dust, and other instrumental effects across the detector.  

\begin{figure}
    \centering
    \includegraphics[width=1.\linewidth]{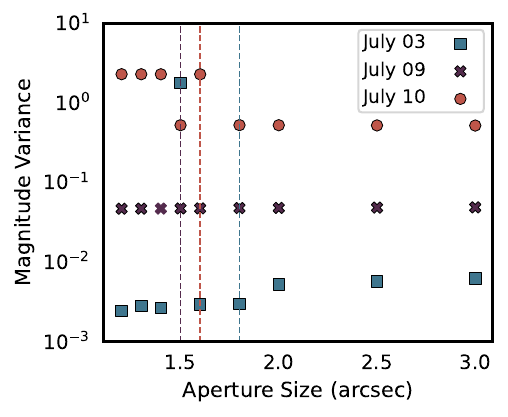}
    \caption{Comparison of the variance of the magnitudes measured as a function of the different aperture sizes used during photometry performed on each set of SOAR observation of 3I/ATLAS. The teal, purple, and orange vertical lines correspond to the aperture size chosen for final photometry of 3I/ATLAS on UT 2025 July 3, 9, and 10, respectively.} 
    \label{fig:ap_test}
\end{figure}

At each stage, the raw bias, flat frames, and science images were first prepared by subtracting and trimming the overscan section of the CCD, a portion of the CCD chip that was not exposed to light. Once the biases were prepared, they were combined into a single combined bias that was subtracted from the other images. The flat frames were similarly calibrated, then combined into a single combined flat frame for each filter. The resulting calibration images that had been reduced and combined were used to produce final reduced science images. The combined bias was subtracted, and finally each of the science images were divided by the combined flat frame that was specific to the filter of the images. 

A precise World Coordinate System (WCS) solution was acquired by submitting the reduced images to \texttt{Astrometry.net} \citep{lang2010}. The \texttt{Astrometry.net} software utilizes a plate solving method in which Right Ascension (RA) and Declination (Dec) coordinates of the image are determined by comparing to stars within field to a database of known stellar positions. The United States Naval Observatory B (USNO-B) \citep{2003AJ....125..984M}, Two Micron All-Sky Survey (2MASS) \citep{2006AJ....131.1163S}, and Gaia Data Release 2 (DR2) \citep{Dotson2024, GaiaCollab2016} catalogs make up the reference catalog used by the \texttt{Astrometry.net} software to determine the WCS solution for a given image.

\subsection{Photometry} \label{subsec:phot}

\begin{figure}
    \centering
    \includegraphics[width=1.\linewidth]{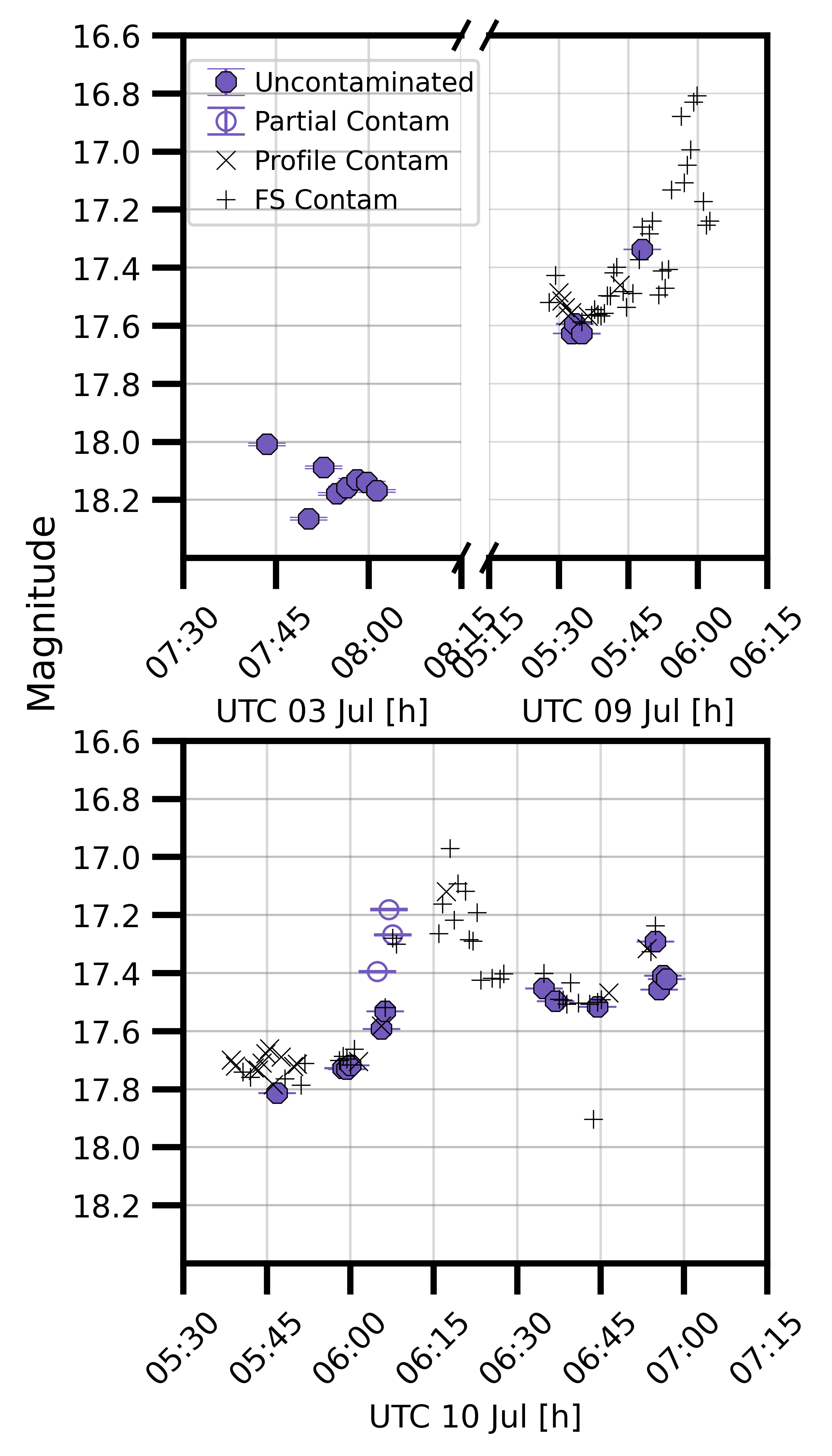}
    \caption{Light curve of 3I/ATLAS including $r'$-band data from the  SOAR telescope observations from UT 2025 July 3, 9, and 10. SOAR data are represented as octagons. The `X' markers indicate points with field star contamination and `+' markers indicate points with radial brightness profile contamination. The open circle markers indicate data that was not identified to be field star or profile contamination, but exhibited unusual brightness variation.}
    \label{fig:soar_lightcurve}
\end{figure}

Photometric analysis of the images of 3I/ATLAS was performed for each night of observations presented in this paper. Our photometry procedure utilized Astropy \texttt{photutils} package to do aperture photometry of 3I/ATLAS.  The aperture was placed on the image by querying the JPL Horizons database for the position of 3I/ATLAS in its orbit at the time of the exposure. The \texttt{CALVIACat} calibration software \citep{michael_kelley_2024_10797842} was used in this procedure to calculate the zeropoint magnitudes for each exposure by comparing photometric values of background stars in the field to the ATLAS ALL-Sky Stellar Reference Catalog (ATLAS-RefCat2) \citep{2018ApJ...867..105T}.

\begin{figure}
    \centering
    \includegraphics[width=1.\linewidth]{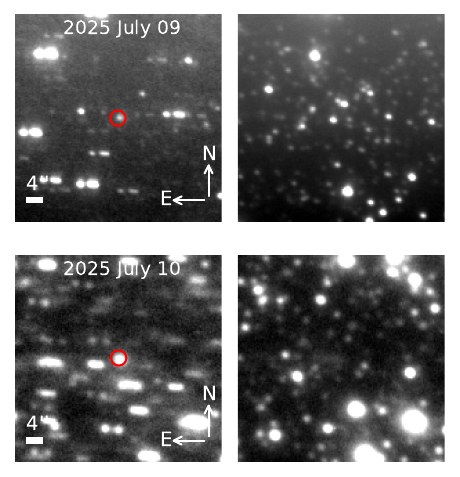}
    \caption{Rotated and stacked composite images of 3I/ATLAS and comparison background stars on UT 2025 July 9 and July 10.} 
    \label{fig:stacks}
\end{figure}

\subsubsection{Aperture Experiments}
 We performed experiments with varying aperture size to test for stellar contamination effects. Specifically, we experimented with aperture sizes ranging from a fraction of to a few times the full width half maximum (FWHM) of the point spread function (PSF) ($1\farcs2$--$3\farcs0$). To this end, we reran all photometric analysis with aperture sizes ranging from $1\farcs2$--$3\farcs0$. We then calculated the standard deviation of the scatter of all photometric points within each night as a function of the aperture radius; this is shown for all nights in Figure \ref{fig:ap_test}. It is  evident that the aperture size does not appear to significantly impact the scatter in the brightness.

Given that there is no clear dependence of the variance of the photometric brightnesses over this dynamic range of aperture size, we somewhat arbitrarily choose the FWHM as the aperture size on each night. The aperture size for the photometric measurements of each night were determined by sampling the FWHM of the target's radial flux profile at various points in the observation. We opted to implement a $1:1$ scaling of the FWHM to determine aperture size to minimize stellar contamination in the aperture given the crowded field on each night. Using that method, we estimated $1\farcs8$, $1\farcs5$, and $1\farcs6$ apertures for UT 2025 July 3, 9, and 10, respectively for the final photometric measurements.

\begin{figure*}
    \centering
    \includegraphics[width=1.\linewidth]{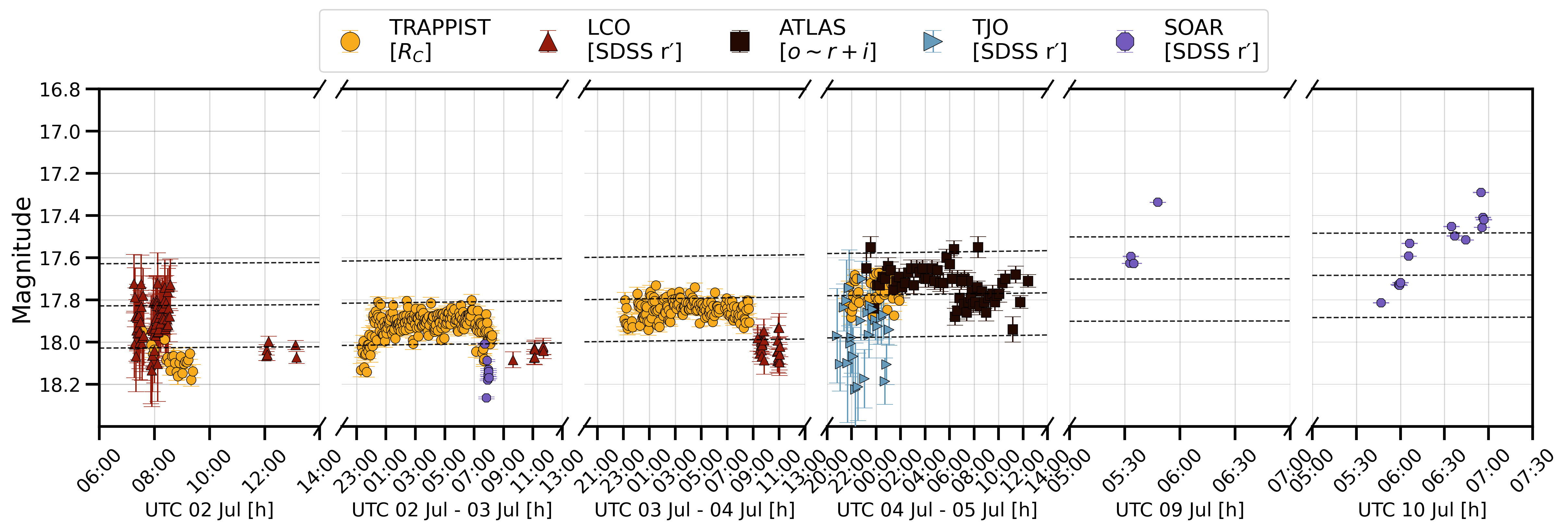}
    \caption{Composite light curve of 3I/ATLAS (adapted from \citet{Seligman2025}) including R\textit{c}-band data from the TRAPPIST Telescopes (-North and -South), \textit{o}-band from the ATLAS Telescopes (Hawai'i, Sutherland, and Canary Islands), and $r'$-band data from the LCO Telescopes, Faulkes Telescopes (-North and -South), Telescope Joan Or\'o, and SOAR Telescope observations from UT 2025 July 3, 9, and 10. Uncontaminated SOAR data are represented as purple octagons. The dotted horizontal lines are lines of constant apparent magnitude given a constant absolute magnitude (with arbitrary zero points) and geocentric distances of 3I/ATLAS (queried from JPL Horizons).}
    \label{fig:comp_lightcurve}
\end{figure*}
\subsubsection{Stellar Contamination Search}
Finally, we performed a search for stellar contamination and rejected photometric points that were significantly contaminated. In Figure \ref{fig:soar_lightcurve} we show the resulting photometric data points, including points that are affected by contamination, from these three nights of SOAR data. Images in which 3I/ATLAS was contaminated were identified by both (i) visual inspection and (ii) assessment of the radial brightness profile compared to the background. On UT 2025 July 10 at 6:00-6:15, $\sim$ 3 data points not flagged by previous contamination detection methods exhibited significant brightness variation. Upon visual inspection, we conclude that this is a result of overall brightening of the field as a result of an artifact of image reduction. These partially contaminated points are marked as open circles in Figure \ref{fig:soar_lightcurve} and are removed from Figure \ref{fig:comp_lightcurve}. Using those methods, photometric data points from UT 2025 July 3 did not show significant contamination.

\subsubsection{Background Subtraction}

Finally, the background level was estimated and subtracted in each image. The background was estimated in each image as the median of a set of pixels with no stellar contamination. That background was then subtracted before converting the counts into a magnitude. 

\subsection{Image Stacking}

To maximize the signal-to-noise and visibility of faint cometary activity, we constructed composite images of the object for each night of data by shifting individual images to align them on the object’s photocenter using linear interpolation and then adding them together. In order to maximize the visibility of faint background stars, we also perform the same procedure aligning on a nearby field star.

In Figure \ref{fig:stacks} we show the stacks of all noncontaminated images. We show stacks from UT 2025 July 9 and UT 2025 July 10 aligned on both 3I/ATLAS and field stars. There is no apparent contamination, which serves as a validation that our uncontaminated subset is correctly identified as such. We also see indications of a coma that is slightly extended approximately to the West, consistent with other published reports from the same time period as these observations \citep{Seligman2025, Opitom2025, Chandler2025, Cordiner2025, Jewitt2025_NOT}. The coma's westward extension indicates it was in the solar direction at the time of these SOAR observations, although we note that later observations indicate that the tail has since switched to the antisolar direction \citep{Tonry2025, Jewitt2025_NOT, Hoogendam2025, Gray2025}. Coma morphologies in the solar direction have been observed in Solar System comets including 19P/Borrelly which was reported to a have a strong sunward facing jet aligned with the spin axis of its nucleus \citep{Farnham2002, Farnham2009}.

\section{Composite Light Curve} \label{sec:comp_lc}

It is evident from Figure \ref{fig:soar_lightcurve} that there was significant scatter in the magnitude of the brightness variations in 3I/ATLAS over UT 2025 July 9 and 10. Moreover, the large brightness increases tend to cluster in time. For example, see the clusters of seemingly sporadic brightness increases at UT July 9 5:53 AM and at UT July 10 6:09 AM.  This alone provides tentative evidence that the seemingly sporadic brightness increases were produced by background stellar contamination. As described in the previous section the visual inspection and PSF comparison revealed the presence of significant contamination in a large subset of our data.  Therefore, each photometric measurement which was contaminated is indicated in Figure \ref{fig:soar_lightcurve}. Observational challenges considered in this work may have affected other near discovery photometric measurements. Future work to consolidate these data sets might provide a more homogeneous data set that corrects for these challenges and contextualizes early photometric data of 3I/ATLAS.

In Figure \ref{fig:comp_lightcurve} we show the extant photometry we obtained with each frame without the contaminated points. It is clear that the brightness increases are entirely caused by  stellar contamination. It should be noted that each data point shown in Figure \ref{fig:comp_lightcurve}  --- including these SOAR photometric data --- have not been corrected for changing observing geometry, including heliocentric and geocentric distances. Given the distances from the first and last set of observations there is an expected brightening of $\sim 0.3$ mag. The variation between each data set seen in the composite light curve in Figure \ref{fig:comp_lightcurve} is due to differing photometric measurement methods (PSF and aperture photometry), observing geometry, zero point selection, selected aperture size, and filter. Table 2 in \citet{Seligman2025} details the different observational and photometric parameters for the TRAPPIST, LCO, ATLAS, and TJO data. For emphasis, these photometric data points are not adjusted to a common filter. This contributes to the brightness variation between different filters seen in Figure \ref{fig:comp_lightcurve}, specifically the SDSS  $r'$ and TRAPPIST R$_c$ data taken contemporaneously on UT 2025 July 3. Given these variations in reduction techniques and filter used, we caution the interpretation of the dotted lines of constant apparent magnitude plotted in Figure \ref{fig:comp_lightcurve}. We add these lines simply to offer a more quantitative assessment of how the brightness measurements of 3I/ATLAS may have evolved in consideration of geocentric distances (queried from JPL Horizons). The mean brightness of 3I/ATLAS on UT 2025 July 3, 9 and 10 from these data points is 18.14, 17.55, and 17.54, respectively.

\section{Discussion} \label{sec: discussion}

In this paper, we presented early SOAR imaging observations of 3I/ATLAS with the Goodman Red Imager on the nights of UT 2025 July 9 and UT 2025 July 10. We added $\sim28$ images to the photometric light curve of 3I/ATLAS from the days immediately following the discovery. The magnitudes of these photometric points are consistent with the previous results presented by \citet{Opitom2025,Seligman2025,Marcos2025,Belyakov2025,Chandler2025}. The mean brightnesses of the uncontaminated photometric data points that we derive are consistent with the brightness of the object upon discovery. However the temporal sparseness of the data and the by-now well-documented strong activity of the object at the time of these observations \citep{Seligman2025,Chandler2025} means that these data are unlikely to provide meaningful constraints on the properties of the nucleus's rotational light curve.

These SOAR observations captured a flat light curve on UT 2025 July 3, consistent with composite light curve measurements reported to date. This implies tentatively that the production of dust --- and perhaps volatiles although that would be inferred assuming some correlation of the dust and gas outburst---  has been somewhat consistent from discovery through the dates when these observations were taken. 
 
It has been noted by \citet{Marcos2025} and \citet{Santana-Ros2025} that the light curve as it develops is presenting a periodic signal. However, due to the relatively small temporal baseline of the new data presented here and the sparseness of it, it would be impossible to differentiate between any periodic signals with these data if they were present. Furthermore, the high levels of sustained activity from the coma likely suppress any noticeable photometric variation caused by the rotation of the nucleus. We explicitly do not speculate on the validity of the periodic signals identified by other authors. Therefore, we do not perform a comprehensive search for periodic signals of activity in the composite light curve. This would be beneficial in future work once more data has been obtained and ideally reduced in a homogeneous manner.  

This work contributes data to the overall secular light curve of 3I/ATLAS. The secular light curve of a comet in general can provide constraints on the composition; for example, outbursting events near relevant sublimation fronts can provide circumstantial evidence for the presence of a given volatile. Moreover, precovery observations of comets can significantly lengthen the arc of a secular light curve and provide information regarding plausible compositions. For example \citet{Ye2020} identified precovery observations of 2I/Borisov in survey data from the Zwicky Transient Facility (ZTF). These archival observations when combined into a secular light curve provided valuable constraints on the nucleus size and volatility of ices driving activity. Another example is provided by \citet{Hui2019} who reported nondetections of 1I/`Oumuamua in Solar and Heliospheric Observatory (SOHO) and Solar TErrestrial RElations Observatory (STEREO)  images. These provided upper limits on both dust and water production when the object was at $\sim0.25$ au. The secular light curve of 3I/ATLAS has already been extended to several months in the past by the LSST precovery observations \citep{Chandler2025} and \textit{TESS} data \citep{Feinstein2025,Martinez-Palomera2025}. These precovery observations indicate that the activity of 3I/ATLAS outwards of 6 au is possibly driven by the sublimation of ices more volatile than H$_2$O. The observations reported here are consistent with the brightness that was reported near discovery, and indicate that this activity has been sustained through July 10. 

Space based follow up observations would allow continued monitoring of cometary activity as 3I/ATLAS moves towards a perihelion of $q\sim 1.36$ au. \citet{Yaginuma2025} demonstrated that a spacecraft sent from Mars could achieve a flyby of 3I/ATLAS with a relatively small $\Delta V$. Therefore, it may be feasible that a space based in situ measurement of our third interstellar visitor will also occur in the upcoming months. \citet{Loeb2025b} demonstrated that a post-perihelion encounter would be feasible if a spacecraft currently at Jupiter was repurposed, such as \textit{Juno}. \cite{Sanchez_2025} demonstrated that a mission like \textit{Comet Interceptor} \citep{Snodgrass2019, Jones2024} would have not been able to reach 3I/ATLAS. 

Given the discovery of 3I/ATLAS --- as well as the spatial number density of interstellar object implied by 1I/`Oumuamua --- all sky surveys should identify more interstellar objects in the near term future. Of course, the NSF-DOE funded Vera C. Rubin Observatory Legacy Survey of Space and Time (LSST) first light was serendipitously almost coincident with the discovery of 3I/ATLAS, which enabled precovery images to be identified during science validation time of Rubin \citep{Chandler2025}. The LSST should identify more interstellar objects at a higher cadence than we have identified them in the previous decade given its increase in search sensitivity
\citep{Moro2009,Cook2016,Engelhardt2017,Hoover2022,Marceta2023a,Dorsey2025}.

\section{Acknowledgments}

We thank the anonymous reviewer for their insightful comments and suggestions that strengthened the scientific content of this manuscript.

T.T.F. and A.Y. acknowledge financial support from NSF grant number AST-2303553. D.Z.S. is supported by an NSF Astronomy and Astrophysics Postdoctoral Fellowship under award AST-2303553. This research award is partially funded by a generous gift of Charles Simonyi to the NSF Division of Astronomical Sciences. The award is made in recognition of significant contributions to Rubin Observatory’s Legacy Survey of Space and Time. 

The SOAR telescope is funded by the National Science Foundation's (NSF) National Optical-Infrared Astronomy Research Laboratory (NOIRLab). SOAR is operated in partnership with Michigan State University (MSU), Minist\'erio de Cience, Tecnologia e Inovaçoes (MCTI) of the Federal Republic of Brazil, and the University of North Carolina at Chapel Hill (UNC). 

We thank Ted Leandro De Almeida for assistance with obtaining data with SOAR. We thank Adina Feinstein, Meir Schochet, and Dennis Bodewits for useful conservations.

\section{Data and Software Availability}
 
All Python scripts used for analysis and reproducing figures will be available on GitHub once published. Larger raw and reduced data files will be available on Zenodo once published.

\textit{Facilities}: Southern Astrophysical Research Telescope (SOAR) 

\textit{Software}: \texttt{astropy} \citep{astropy:2013, astropy:2018, astropy:2022}, \texttt{CALVIACat} \citep{michael_kelley_2024_10797842}, \texttt{matplotlib} \citep{Hunter:2007}, \texttt{numpy} \citep{harris2020array}, \texttt{scipy} \citep{2020SciPy-NMeth}, \texttt{Astrometry.net} \citep{lang2010}.

\bibliographystyle{mnras}
\bibliography{main}

\bsp	
\label{lastpage}
\end{document}